\documentclass[traditabstract]{aa}
\usepackage[varg]{txfonts}

\usepackage{graphicx}
\usepackage{natbib}
\usepackage{amsmath}
\usepackage{amssymb}
\usepackage{url}

\newcommand{\kms}[0]{\mbox{km\,s$^{-1}$}}

\newcommand{\ergs}[0]{\mbox{erg\,s$^{-1}$}}
\newcommand{\ergcms}[0]{\mbox{erg\,cm$^{-2}$\,s$^{-1}$}}
\newcommand{\cv}[0]{CVSO~30}
\newcommand{\cvb}[0]{CVSO~30\,b}
\newcommand{\cvc}[0]{CVSO~30\,c}
\newcommand{\xmm}[0]{XMM-\textit{Newton}}
\newcommand{\chan}[0]{Chandra}
\newcommand{\cts}[0]{\mbox{ct\,s$^{-1}$}}
\newcommand{\ctks}[0]{\mbox{ct\,ks$^{-1}$}}
\newcommand{\teff}[0]{\mbox{T$_{\mathrm{eff}}$}}
\newcommand{\mj}[0]{\mbox{M$_{\mathrm{J}}$}}
\newcommand{\myr}[0]{Myr}

\begin{document}

\title{X-ray emission in the enigmatic \cv\ system}
\author{S. Czesla\inst{1} \and P.C. Schneider\inst{1} \and M.
Salz\inst{1} \and T. Klocov\'a\inst{2} \and T. O. B. Schmidt\inst{1} \and J. H. M. M. Schmitt\inst{1}}
\institute{Hamburger Sternwarte, Universit\"at Hamburg, Gojenbergsweg 112, 21029
Hamburg, Germany \and Astronomical Institute ASCR, Fricova 298, 25165 Ondrejov, Czech Republic}
\date{Received ... / Accepted ... }

\abstract{\cv\ is a young, active, weak-line T Tauri star; it possibly hosts the only known
planetary system with both a transiting hot-Jupiter and a cold-Jupiter candidate (\cvb\ and \cvc). We
analyzed archival ROSAT, \chan, and \xmm\ data to study the coronal emission in the system.
According to our modeling, \cv\ shows a quiescent X-ray
luminosity of $\approx 8\times 10^{29}$~\ergs. The X-ray absorbing column is consistent with interstellar absorption.
\xmm\ observed a flare, during which a transit of the candidate \cvb\
was expected, but no significant transit-induced variation in the X-ray flux is detectable. While the hot-Jupiter candidate
\cvb\ has continuously been undergoing mass loss powered by the high-energy irradiation, we conclude that its evaporation
lifetime is considerably longer than the estimated stellar age of $2.6$~\myr.
}

\maketitle

\section{Introduction}

\cv\ is classified as a young weak-line T Tauri star (WTTS) in the Orion OB~1a association \citep[][and Table~\ref{tab:pars}]{Briceno2005}.
At an age of $2.6$~\myr, \cv\ may host
one of the youngest planetary systems known to date and the only one comprising both a
hot- and a cold-Jupiter candidate \citep[\cvb\ and \cvc,][]{Eyken2012, Schmidt2016}.
Both planetary candidates remain controversial to date \citep[][]{Yu2015, Onitsuka2017, Lee2018}.

The presence of the transiting  hot-Jupiter candidate \cvb\ was announced in 2012 when
\citeauthor{Eyken2012} discovered recurring photometric fading
events\footnote{\citet{Eyken2012} actually dubbed them transits;
here we adopt the nomenclature also used by \citet{Yu2015}.} with a
period of $\approx 0.44$~d.
These fading events are superimposed on a
pronounced pattern of activity-induced rotational variability with
a close (or identical) period \citep[see also][]{Koen2015}.
 \citet{Eyken2012} suggested that the fading events may be attributable to transits by
a hot Jupiter with a mass of
M$_\mathrm{p} < 5.5\pm1.4$~\mj.
However, the events show changes in shape, and occasionally even
vanish entirely \citep{Eyken2012, Ciardi2015, Yu2015}. 
To explain this behavior in the context of a planet,
orbital precession was put forward by \citet{Eyken2012} and
detailed studies with theoretical solutions were presented by
\citet{Barnes2013} and \citet{Kamiaka2015}. Some of their results along with other
relevant system parameters are given in Table~\ref{tab:pars}.
Further support for the planetary hypothesis comes from the analysis of variable, asymmetric H$\alpha$ emission 
by \citet{JohnsKrull2016}, who conclude that their observation is  consistent with the existence
of an evaporating planet \cvb.

Nevertheless,
the planetary hypothesis does not remain uncontested. Having juxtaposed a number
of alternatives to explain the fading events, \citet{Yu2015} concluded that
an occulted hotspot model gains the most support from the considered data. The authors found the
existence of a gas giant \cvb\ particularly difficult to reconcile with the lack of an observable
occultation and phase shifts in the timing of the fading events.
In addition, 
the detection of color-dependent fading events by \citet{Onitsuka2017} and \citet{Yu2015} is  at odds
with a planetary origin.
Without digressing into more minute details, we confirm that 
the origin of the fading events remains disputed.
  
Any material in the vicinity of a young, active star like \cv\
is immersed in strong activity-induced extreme ultraviolet and
X-ray emission \citep[e.g.,][]{Stelzer2001}, frequently intensified by
flaring \citep[e.g.,][]{Favata2005}. 
In a planetary atmosphere, such high-energy
irradiation is absorbed in the upper layers, which expand and drive off mass in the
form of a planetary wind \citep[e.g.,][]{Salz2016, Salz2016b}.
The high-energy radiation thus plays a decisive role in shaping the
environments of young stars, and 
the point of this paper is to deduce its characteristic properties in the \cv\ system.

\begin{table}
\caption{Parameters of \cv\ and hot-Jupiter candidate \cvb\ used in this study.
\label{tab:pars}}
\begin{tabular}{l l l} \hline \hline
  Parameter & Value & Source\tablefootmark{a} \\
  \hline
  \teff\ [K] & 3470 & B \\
  Spectral type & M3 & B \\
  Age [\myr] & 2.6 & B \\
  v$\sin(i)$ [\kms] & $80\pm8.1$ & E \\
  Distance [pc] & $349\pm 9$ & G \\
  M$_{\rm S}$ [M$_{\odot}$] & 0.44 & Ba \\
  R$_{\rm S}$ [R$_{\odot}$] & $1.03$ & Ba \\
  \hline
  \multicolumn{3}{c}{Planetary candidate \cvb} \\
  T$_{0}$ [HJD] & $2455543.9402 \pm 0.0008$ & E \\
  P$_{\mathrm{orb}}$ [d] & $0.448413 \pm 0.00004$ & E \\
  R$_{\rm P}$ [R$_{\rm J}$] & $1.68\pm 0.07$ & Ba \\
  M$_{\rm P}$ [M$_{\rm J}$] & $3.6\pm 0.3$ & Ba \\
  \hline
\end{tabular}
\tablefoot{\tablefoottext{a}{B: \citet{Briceno2005}}; E: \citet{Eyken2012}; G: Gaia DR2 \citep{Gaia2016, Gaia2018};
Ba: \citet{Barnes2013} (solution for M$_{\rm S} = 0.44$~M$_{\odot}$)}
\end{table}

\section{Observations and data analysis}

In the following, we analyze
archival X-ray data from the \xmm, \chan, and ROSAT X-ray satellites, which
all serendipitously observed \cv.

\subsection{\xmm\ data}
\label{sec:xmmdata}
The \xmm\ satellite is equipped with three X-ray telescopes \citep{Jansen2001}. At their focal plane, there are  three
CCD cameras provided by the European Photon Imaging Camera (EPIC) consortium.
The assembly consists of two metal oxide semi-conductor (MOS) CCD arrays and one array of so-called pn-CCDs
\citep[][]{Stueder2001, Turner2001}. The fields of view of the pn  and MOS cameras largely overlap,
and they are usually operated
simultaneously. All of them record the position, energy, and arrival time
of individual X-ray photon events.

\cv\ was observed by \xmm\ during a  
$63.9$~ks observation nominally targeted at the
25~Orionis cluster carried out in March 2009 (Obs~ID 0554610101). We reduced
the data using \xmm's Science Analysis Software (SAS) version 13.5 using standard data
reduction recipes\footnote{
\url{http://xmm.esac.esa.int/sas/current/documentation/threads/}}, where not
stated otherwise.

\begin{figure}[h]
  \includegraphics[width=0.49\textwidth]{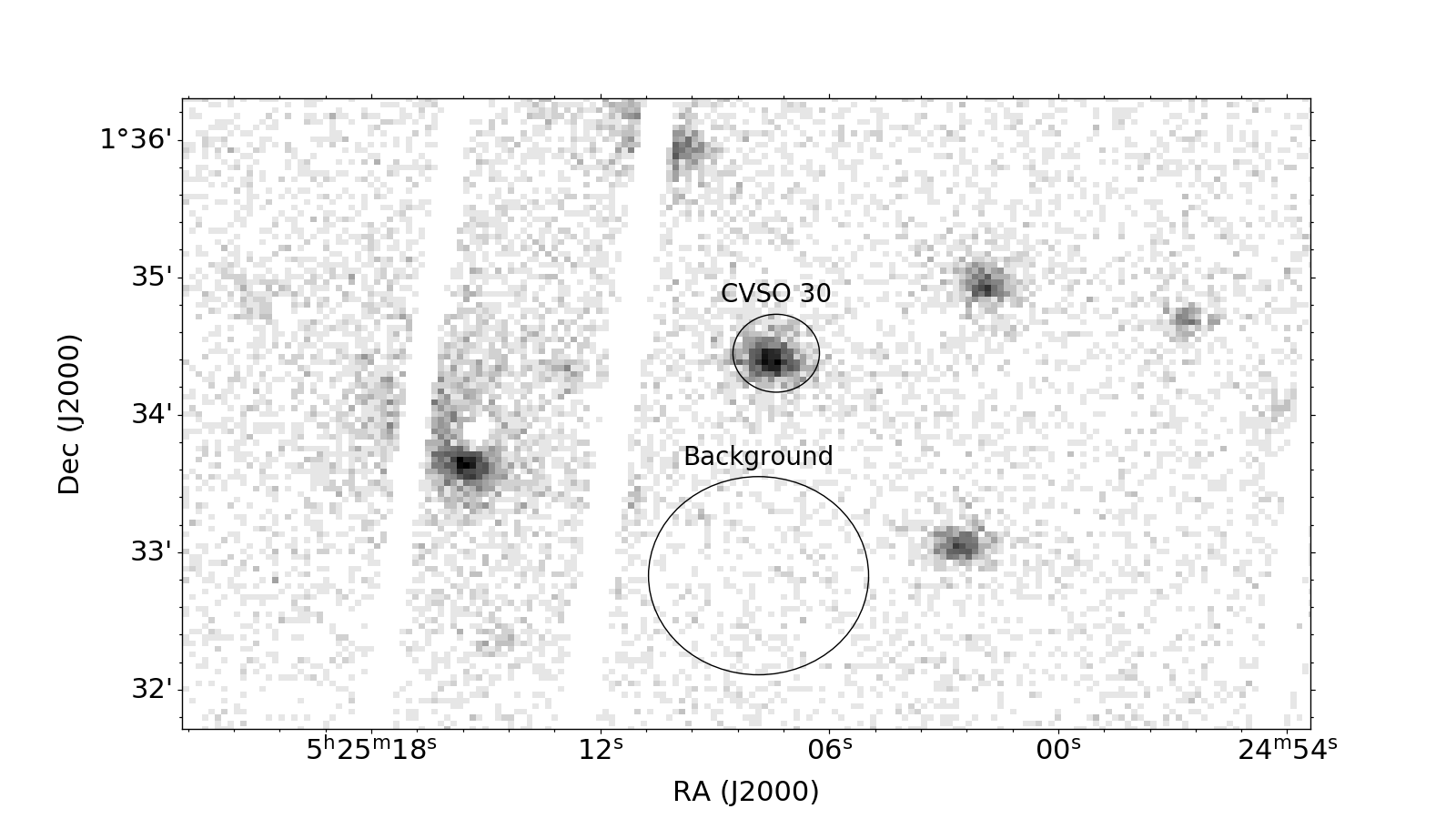} \\
  \caption{\xmm\ pn image images of \cv\ ($0.5-2$~keV band).
  The circles indicate our source and background extraction regions.
  \label{fig:CVSO30XIM}}
\end{figure}

In Fig.~\ref{fig:CVSO30XIM} we show the
$0.5-2$~keV band pn~image of the surroundings of \cv\ along with our
circular source and background extraction regions. The source region has a radius of $17$~arcsec and covers
about $80$\% of the photons on the pn chip (at $1.5$~keV).
There is a clear, point-like X-ray source at the position of \cv\
without distinguishable contamination from nearby sources within at least one~arcminute. 

During the observation
the pn data show a slightly elevated but quite constant
level of high-energy background ($10-12$~keV).
The highest observed pn background count rate is $0.56$~\cts,
which is not too high above the recommended cutoff level of $0.4$~\cts.
Because \cv\ is neither an extended nor particularly weak X-ray source, we expect no
serious background interference in the analysis and, consequently,  
applied no background filtering for the pn data.
\cv\ was also observed with the MOS~2 camera, where high-energy background was no issue.
The MOS~1 camera missed the target because it was positioned on CCD no. 6,
which remained switched off during the observation.

\subsubsection{X-ray photometry}
\label{sec:xmmPhotometry}
\begin{figure}[h]
  \includegraphics[angle=0,width=0.49\textwidth]{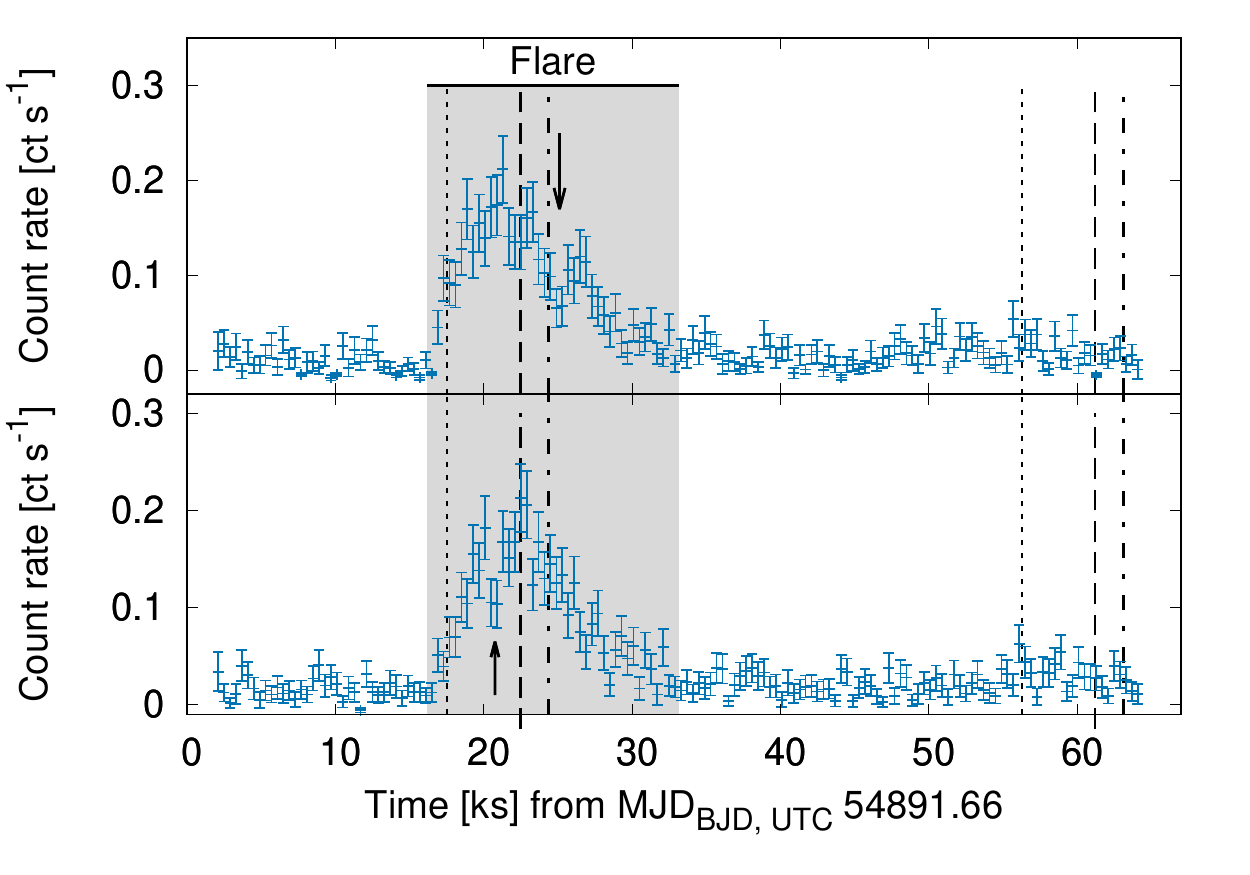}
  \caption{Combined background-subtracted MOS~2 and pn light curves of \cv\
  with a binning of $400$~s. The bottom panel shows the soft-band light curve
  ($0.1-1$~keV) and the top panel the hard-band ($1-9$~keV) counterpart. 
  The adopted flare period is indicated by  gray shading. Vertical lines 
  indicate the times of fading events according to different ephemerides: 
  the linear and quadratic ephemerides given
  by \citet{Yu2015} indicated by dash-dotted and dotted lines,  and the ephemerides by
  \citet{Raetz2016} indicated by dashed lines. Two arrows indicate the potential
  dips discussed in Sect.~\ref{sec:XrayEclipse}.
  \label{fig:pnlc}}
\end{figure}

In Fig.~\ref{fig:pnlc} we show the
combined pn and MOS~2 X-ray light curve of \cv, generated with the SAS task \texttt{epiclccorr}, which
also applies a number of corrections\footnote{\url{https://heasarc.gsfc.nasa.gov/docs/xmm/sas/help/epiclccorr/node3.html}}.
The temporal binning is $400$~s.
The top panel shows the hard-band ($1-9$~keV) light curve and
the bottom panel its soft-band ($0.1-1$~keV) counterpart. The X-ray light
curve shows a strong flare during which the count rate rises by about
an order of magnitude compared to the quiescent level; this flare was
previously  reported by \citet{Franciosini2011}, but no detailed analysis
was carried out.

In Fig.~\ref{fig:pnlc} we also show our definition of the flaring period, which
starts at MJD~$54\,891.847$ and lasts for $17$~ks.
Given that the stellar rotation period is almost
identical to or slightly larger than the planetary orbital period of $39.7$~ks \citep[][]{Eyken2012, Raetz2016},
the flare lasts for
essentially half a stellar rotation period. Even so, the flare light curve shows
no evidence of an occultation of the flare region by the star. As the flaring material is thought to be optically
thin,  no strong line-of-sight effects related to the change in viewing geometry caused by rotation are expected.

In addition, Fig.~\ref{fig:pnlc}
shows the times of the fading events according to both the linear and quadratic
ephemerides given by \citet{Yu2015} and the ephemerides found by \citet{Raetz2016}. In the nomenclature used by \citet{Yu2015},
the covered epochs are $-690$ and $-689$\footnote{These corresponds to epochs $-1453$ and $-1452$
according to the ephemerides given by \citet{Eyken2012} and \citet{Raetz2016}.}.  The ephemerides given by
\citet{Eyken2012} show fading event times preceding those predicted by \citet{Raetz2016} by about $2$~ks.
The formal (statistical) uncertainty of the fading event times
is $0.6$~ks in the case of \citet{Yu2015} and $0.1$~ks in the case of \citet{Raetz2016}.
As the earliest photometric data published by \citet{Eyken2012} were obtained in Dec. 2009 (i.e.,
about six months after the \xmm\ observations), all calculations of the fading event times require extrapolation.
Therefore, we argue that the difference between these different predicted times gives an impression of the true
uncertainty involved. Nonetheless, all considered ephemerides predict that
a fading event may have occurred sometime during the flare.

As may be expected in such an active star, the out-of-flare X-ray count
rate is not strictly constant.
While the mean combined pn and MOS~2 pre-flare count rate is
\mbox{$27\pm 2$~\ctks}, its post-flare equivalent is $40\pm 2$~\ctks, which
may be attributable to intrinsic coronal variability or rotational modulation.
Nevertheless, we treat the pre- and post-flare periods as a single
quiescent period in our analysis.

\subsubsection{Spectral analysis}
\label{sec:spectralXray}
\begin{figure}[h]
  \includegraphics[width=0.49\textwidth]{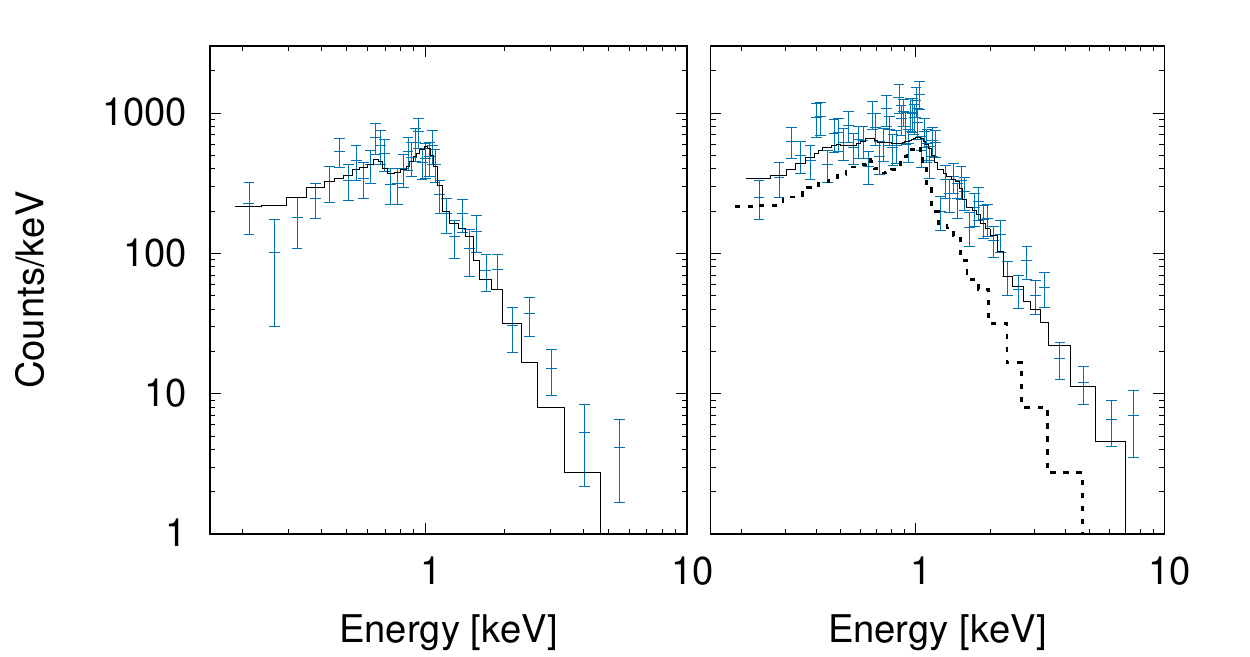}
  \caption{\xmm\ EPIC-pn spectrum and best-fit model (solid line) during
  quiescence (left) and flare (right) periods. For reference, we also show the
  quiescent spectrum in the flare panel (dotted line).
  \label{fig:xrayspec}}
\end{figure}

We carried out a spectral analysis for the quiescent and flare periods as
defined above using XSPEC version $12.9.1$ \citep{Arnaud1996}.
Before the spectral fit, we grouped the observed spectra so that each spectral energy bin contains 15 counts.
To model the spectrum
during the quiescent period, we used a two-temperature model
computed with the
Astrophysical Plasma Emission Code (APEC) for optically thin plasmas in collisional ionization equilibrium
\citep[][]{Smith2001}\footnote{See also \url{http://atomdb.org/}}.
The elemental abundances were all combined into
a single scaling parameter. In modeling the  average  flare spectrum, we assumed
that the contribution from the quiescent phase remained constant
and added a third thermal component with independent abundances to describe the
flare-induced spectral change. The resulting parameters are given in
Table~\ref{tab:fitparams} and the EPIC-pn quiescence and flare spectra along
with the best-fit models are shown in Fig.~\ref{fig:xrayspec}.

\begin{table}
  \centering
  \caption{Results of spectral modeling of the \xmm\ and \chan\ data along with
  $68$\% confidence intervals. 
  \label{tab:fitparams}}
  \begin{tabular}[h]{l l} \hline\hline
  Parameter & Value \\ \hline
  \multicolumn{2}{c}{\xmm\ observation} \\
  N$_H$ [$10^{22}$~cm$^{-2}$] & $0.029_{-0.01}^{+0.01}$ \\
  \multicolumn{2}{c}{Quiescent components} \\
  T$_1$ [keV] & $0.23_{-0.06}^{+0.09}$ \\
  EM$_1$\tablefootmark{c} [$10^{52}$~cm$^{-3}$] & $2.5_{-1.1}^{+1.1}$ \\
  T$_2$ [keV] & $1.24_{-0.04}^{+0.06}$ \\
  EM$_2$\tablefootmark{c} [$10^{52}$~cm$^{-3}$] & $6.8_{-1.3}^{+1.3}$ \\
  Ab$_{1,2}$\tablefootmark{c} & $0.21_{-0.07}^{+0.09}$ \\
  L$_{\mathrm{X,quiescence}}$\tablefootmark{a} [$10^{29}$~\ergs] & 
$8.0_{-0.8}^{+0.8}$
  \\
  \multicolumn{2}{c}{Flare component} \\
  T$_3$ [keV] & $2.4_{-0.4}^{+0.3}$ \\
  Ab$_3$\tablefootmark{c} & $0.12_{-0.09}^{+0.10}$ \\
  EM$_3$\tablefootmark{c} [$10^{52}$~cm$^{-3}$] & $37_{-4}^{+4}$ \\ 
  L$_{\mathrm{X,flare}}$\tablefootmark{a,b} [$10^{29}$~\ergs] &
  $47.2_{-3.3}^{+3.3}$ \\
  \hline
  \multicolumn{2}{c}{\chan\ observations} \\
  \multicolumn{2}{c}{Obs ID 8573\;\;\;(Jan. 8, 2008)} \\
  T [keV] & $1.31_{-0.3}^{+0.4}$ \\
  EM\tablefootmark{c} [$10^{52}$ cm$^{-3}$] & $9.1_{-1.8}^{+1.8}$ \\
  L$_{\mathrm{X}}$\tablefootmark{a,b} [$10^{29}$~\ergs] & $8.8_{-1.5}^{+1.5}$ \\
  \multicolumn{2}{c}{Obs ID 8572\;\;\;(Jan. 13, 2008)} \\
  T [keV] & $0.93_{-0.2}^{+0.3}$ \\
  EM\tablefootmark{c} [$10^{52}$ cm$^{-3}$] & $3.2_{-1.1}^{+1.1}$ \\
  L$_{\mathrm{X}}$\tablefootmark{a} [$10^{29}$~\ergs] & $3.4_{-1.0}^{+1.0}$ 
\\ \hline
  \end{tabular}
  \tablefoot{
  \tablefoottext{a}{Unabsorbed, $0.3-9.0$~keV band};
  \tablefoottext{b}{Includes the quiescent contribution.}
  \tablefoottext{c}{Abundance (Ab) and emission measure (EM)}
  }
\end{table}

According to our modeling
the coronal abundances of \cv\ are subsolar with respect to the solar (photospheric) abundances
given by \citet{Anders1989}.
During the flare the spectrum hardens as expected.
The average flare X-ray luminosity, L$_{\mathrm{X,flare}}$, listed in
Table~\ref{tab:fitparams} includes the quiescent contribution. Subtracting the
quiescent level, we estimate that the flare released a total of
\begin{equation}
(47.2-8.0)\times 10^{29}\mbox{~\ergs} \times 17\mbox{~ks}
=6.8\times 10^{34}\mbox{~erg}
\end{equation}
in the $0.3-9$~keV band. While this is certainly a strong flare compared
to flares  observed on CN~Leo, for example \citep[][]{Schmitt2008}, it still releases
two to three orders of magnitude less energy in X-rays than the strongest
flares observed on young stars of comparable age in Orion \citep{Favata2005}.

\subsection{\chan\ data}
\chan\ observed \cv\ twice on January 8 and   13, 2008 (Obs. IDs 8573 and
8572). The observations lasted for about $10$~ks each and were carried out with the
same setup. The observations were used in the study by \citet{Ingleby2011},
but no analysis of the specific source \cv\ is presented there.
In both observations,
\cv\ is found far from the aim point on the ACIS-I detector
where \chan's point spread function (PSF) is much broader
than for on-axis sources. Having inspected nearby stronger X-ray sources, we
opted for a circular extraction region with a radius of $25$~arcsec placed on
the nominal
source position to carry out the analysis.

In both of the \chan\ images an X-ray source at the location of \cv\ is clearly
visible. In particular, we find $44.9$ and $18.4$ excess photons in the source
regions in the observations 8573 and 8572.
We used CIAO version 4.7 with CALDB version 4.6.5 to obtain
spectra and light curves using standard analysis
threads\footnote{\url{http://cxc.harvard.edu/ciao/threads/index.html}}.

Both \chan\ light curves of \cv\ are compatible with a constant source count
rate without major flares or strong gradients. Therefore, we proceeded to obtain
spectra, which we grouped to five counts per spectral bin. The spectra were
modeled using an absorbed single-component thermal plasma model using XSPEC.
Because of the low number of counts, we specifically assumed Poisson-distributed 
source and background rates in our analysis (\texttt{cstat} option in XSPEC). 
As the interstellar hydrogen absorption column and the coronal
elemental abundances are not constrained by the \chan\ data, we fixed
them to the values derived from the \xmm\ observation.
Our fit results are also listed in
Table~\ref{tab:fitparams}.

The temperature, emission measure, and also X-ray luminosity derived from the
\chan\ observation on January 8 are compatible with the values found during the
quiescent phase observed by \xmm. For the observation on January 13 our modeling yields
a source fainter by about a factor of two and also somewhat cooler than
on January 8.
Both \chan\ observations are relatively short and, factoring in the usual degree of variability in
the X-ray flux of young stars, we consider the results consistent
with the quiescent phase observed by \xmm.

\subsection{ROSAT data}

ROSAT observed the \cv\ system for $23$~ks with the High Resolution Imager (HRI)
in March 1994 (Rosat Observation Request 201392h).
Although \cv\ was located rather far from the aim point,
we detected a clear source with a best-fit position $8$~arcsec from
the nominal position of \cv, which is quite typical for sources
observed in ROSAT~HRI pointings \citep[e.g.,][]{Sasaki2000}.
In particular, we found an excess of $30\pm8.5$~cts corresponding to an HRI
count rate of $1.4\pm0.4$~\ctks.

Using
\texttt{WebPIMMS}\footnote{\url{http://heasarc.gsfc.nasa.gov/cgi-bin/Tools/w3pimms/w3pimms.pl}},
we derived a count--energy conversion factor of $3.67\times
10^{-11}$~erg\,cm$^{-2}$\,s$^{-1}$\,ct$^{-1}$ ($0.3-9.0$~keV band) assuming a
$1.2$~keV thermal model with 20\% solar abundances (cf. Sect.~\ref{sec:xmmdata})
and an absorption column of $2.9\times 10^{20}$~cm$^{-2}$.
This yields an estimate of $(6.7\pm1.9)\times 10^{29}$~\ergs\ for
the unabsorbed X-ray luminosity of the target, which is
consistent with the quiescent luminosities obtained from the
analysis of the \xmm\ and \chan\ data.

\section{Discussion}

\subsection{Quiescent X-ray spectrum}

From our analysis of the ROSAT, \xmm, and \chan\ data, we deduced consistent
X-ray luminosities of about $8 \times 10^{29}$~\ergs\ for \cv.
This level of quiescent X-ray emission is compatible with the distribution of
X-ray luminosities of WTTSs with similar mass in Orion and M16, whose age is
about $1$~\myr\ \citep[][]{Getman2005, Guarcello2012}. Formally, the relation
between mass and X-ray luminosity given by \citet{Guarcello2012} (their Eq.~2)
yields an estimate of $2.9 \times 10^{30}$~\ergs\ for a WTTS (i.e., a class~III
object) with a mass of $0.4$~M$_{\odot}$, where the scatter of the
distribution amounts to about two orders of magnitude.
The value is also compatible with the mean X-ray luminosity of $(10 \pm
2.5)\times 10^{29}$~\ergs\ of fast-rotating ($v\sin(i) > 22$~km\,s$^{-1}$) WTTSs
in the Taurus-Auriga-Perseus region \citep[][]{Stelzer2001}.

Our finding of subsolar coronal abundances in \cv\ with respect to the solar photosphere (Table~\ref{tab:fitparams})
is not uncommon for highly active stars \citep[e.g.,][]{Scelsi2007}. We are not aware of an estimate
of the photospheric abundances in \cv\ that can be used for comparison. The relation between
coronal and photospheric abundances can be rather complex in the Sun and other stars
\citep[e.g.,][]{Feldman1992, Laming1996, Drake1997, Guedel2009}. 

Our analysis indicates that the quiescent level of X-ray emission of \cv\
remained stable to within a factor of about two for at least $15$~yr, i.e.,
the time between the ROSAT and \xmm\ observations. This finding is in line
with the results reported by \citet{Telleschi2007}, who find a similar variation between
ROSAT and \xmm\ observations in their study of nine T Tauri stars.
 Other highly active stars such as the $50$~Myr old AB~Dor and
the RS~CVn binaries AR~Lac and HR~1099 also show  similarly constant levels of
X-ray emission, disregarding
strong short-term variability due to flares \citep{Lalitha2013, Drake2014, Perdelwitz2018}.

With a bolometric luminosity of $0.25$~L$_{\odot}$ \citep{Eyken2012},
a quiescent X-ray luminosity of $8\times 10^{29}$~\ergs\
yields a $\log_{10}(L_{\mathrm{X}}/L_{\mathrm{bol}})$ ratio of $-3.1$. This 
puts \cv\ close to the
saturation limit of $-3$ \citep{Pizzolato2003}, which is consistent with it
being a young pre-main-sequence star. Adopting the radius of $1.03$~R$_{\odot}$
given by \citet{Barnes2013},
we derive a value of $\log_{10}(F_X\;\mbox{[erg\,cm$^{-2}$\,s$^{-1}$]}) = 7.1$
for the X-ray surface flux.
This value is not unusual for fast-rotating T Tauri stars \citep{Stelzer2001}
or the most active K- and M-type stars in the solar neighborhood
\citep[][Fig.~8]{Schmitt1997}.

\subsection{Absorption column density}

While extinction at photon energies in excess of $0.53$~keV, i.e., the
K-shell edge of oxygen, is dominated by heavy
elements irrespective of their chemical state, optical
extinction is mainly caused by dust grains \citep[e.g.,][]{Predehl1995,
Guever2009, Wilms2000, McJunkin2014}. The depth of the hydrogen column and the amount of
optical extinction are closely correlated in the interstellar medium so that $N_H [\mbox{cm$^{-2}$}] \sim A_V
[\mbox{mag}]$. However, published constants of proportionality differ by 
up to about $25$\% \citep[see][]{Guever2009}.

Adopting the value of $1.79\times
10^{21}$~cm$^{-2}$ reported by \citet{Predehl1995}, we converted the 
hydrogen column density of
$\approx 2.9\times 10^{20}$~cm$^{-2}$ deduced from the \xmm\ data into an
optical extinction coefficient, $A_V$, of $0.16 \pm 0.06$~mag. This value
is compatible with the number of $0.12$~mag given by \citet{Briceno2005} for
\cv. The properties of the absorbing material are  therefore  consistent with
those of the interstellar medium; in particular, no significant excess of gas or dust 
extinction is measured. Assuming an average density of one particle per cm$^3$ along the
line of sight \citep{Welsh2010} yields an estimate of $10^{21}$~cm$^{-2}$ for the
interstellar column density toward \cv.
Therefore, is seems plausible that the observed absorption is largely interstellar in origin.  

\subsection{Any X-ray fading events?}
\label{sec:XrayEclipse}
While the predictions for the timing of the fading events differ depending on the
assumed ephemerides (see Sect.~\ref{sec:xmmPhotometry}), it appears plausible that
the flare covers one such event. Given the temporary disappearance of the optical fading events,
we cannot be sure that any took place at all in this epoch however. Nonetheless,
we scrutinized the light curves for any X-ray variability potentially associated
with optical fading.

It is clear that the transit of a planetary disk causing a small optical depression would be impossible to detect in X-rays with the data at hand unless a compact flaring
region was eclipsed.
For example, \citet{Briggs2003} 
report a $60$\% drop in the X-ray count rate lasting for a total of about $1.4$~ks
during an X-ray flare on the K3V-type Pleiades member star H~II~1100, which  they argue  
might plausibly be caused by a hot Jupiter eclipsing the flaring region.
However, they  caution that 
the passage of cool and dense prominence material across the flaring region
\citep[see also][]{Haisch1983} or a peculiar superposition of flare events also
provide viable explanations. In any case,
the transit of a more extended cloud of material, whether associated with a
planet or not, might be more easily detectable.

About $20$~ks into the \xmm\ observation (Fig.~\ref{fig:pnlc}),
the soft-band count rate drops by $\approx 35$\,\% for
$\approx 800$~s, i.e., two data points. This timescale is consistent with the occultation of a point-like source
by the hypothesized planetary body of \cvb.
However, the lack of an associated depression
in the hard-band light curve makes an eclipse by a solid planetary disk unlikely.
A temporary elevation
in the absorption column by $10^{21}$~cm$^{-2}$ (e.g., caused by the
transit of an extended planetary atmosphere) produces a count rate drop of
$60$\,\% in the soft band with only a $10$\,\% drop in the hard band. Such a scenario would thus
be compatible with
the observation.
By means of a Student's t-test, we find that the mean of the two data points in question differs from
that of the sample consisting of the two preceding and following points at the $2$\,\%
significance level. As the ephemerides are uncertain, however, we count about  
ten similar opportunities during the flare where such an occultation may have happened and be interpreted in
the same way. Therefore, the chance to obtain such a result by chance
is about $20$\,\%, and we 
conclude that the data remain insufficient to produce a significant result.

The hard-band light curve shows an apparent depression about 25~ks into the \xmm\
observation (Fig.~\ref{fig:pnlc}). This variation has no counterpart in the soft-band light curve, however, and might be
more easily explained by the superposition of two flares or secondary heating rather than absorption or occultation of
flaring material.

\subsection{Planetary irradiation and mass loss}
\label{sec:MassLoss}

\begin{figure}
  \includegraphics[width=0.48\textwidth]{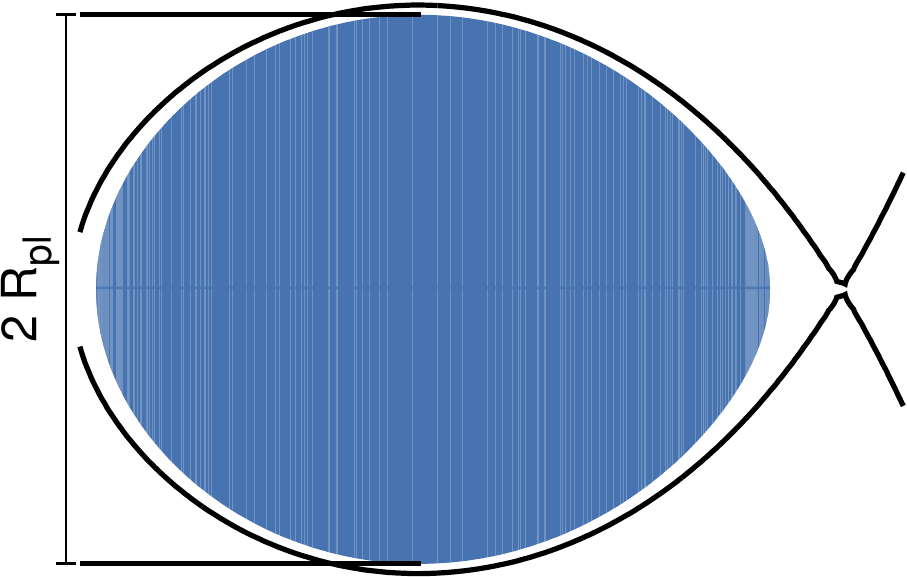}
  \caption{Roche lobe (solid line) and extent (blue shade) of planetary candidate \cvb\
                    in the orbital plane (0.44~M$_{\odot}$, 3.9~M$_{\mathrm{J}}$,
           and 1.61~R$_{\mathrm{J}}$).
           }
  \label{fig:rochelobe}
\end{figure}

Having measured the stellar X-ray flux, we 
now estimate the expected mass-loss rate of the candidate hot Jupiter \cvb\ to
see whether such a configuration is physically
plausible.
A number of values have been proposed for the mass of \cvb\ (e.g.,  \citealt{Barnes2013}, \citealt{Ciardi2015},
or \citealt{JohnsKrull2016}). These range from approximately one to seven Jovian masses.
We  assume here 
a stellar mass of 0.44~M$_{\odot}$, a planetary mass of 3.9~M$_{\mathrm{J}}$,
and a planetary radius of 1.61~R$_{\mathrm{J}}$. These values are based on those given in Table~\ref{tab:pars}.
To remain on the conservative side regarding the planetary mass-loss rate,
we increased the nominal mass estimate in Table~\ref{tab:pars} by its uncertainty and decreased the
radius likewise. Both choices increase the assumed planetary density and
inhibit mass loss.
Kepler's third law then demands
a semimajor axis, $a$, of $0.00874$~AU.
The resulting Roche-lobe geometry is shown in Fig.~\ref{fig:rochelobe}. The shape of the planetary
candidate \cvb\ is determined by an equipotential surface, the level of which is given by the radius derived
from transit modeling. The result clearly demonstrates
the expected deviation from spherical geometry.

Energy-limited escape is based on the assumption that
the complete energy input obtained through high-energy irradiation is used to
lift planetary material from the surface to the Roche lobe height \citep{Watson1981}.
Substituting a quiescent X-ray luminosity of $8\times 10^{29}$~\ergs\ yields an
irradiation of $3.7\times 10^6$~\ergcms\ at the distance of the planet.
Curiously, the level of surface
irradiation varies by about $10$\,\% because the substellar point on the planetary
surface is noticeably closer to the star than the terminator is 
as the orbital semimajor axis corresponds to a mere $18$~Jupiter
radii. 

Using the relation between the X-ray ($5-124$~\AA) and extreme ultraviolet (EUV,
$124-912$~\AA) stellar surface flux presented by
\citet[][Eq.~2b]{Chadney2015}, we estimate a value of $1.2\times 10^{30}$~\ergs\ for the
combined stellar EUV and X-ray (XUV) luminosity of \cv.
The resulting XUV flux at the planetary orbital distance
then becomes $F_{XUV} = 5.4\times 10^6$~\ergcms.
The energy-limited planetary mass-loss rate is given by
\begin{equation}
  \dot{M} = \frac{3 \eta F_{XUV}}{4 K G \rho_{\mathrm{pl}}} \; ,
\end{equation}
where $\eta$ is the heating efficiency, $G$ is the gravitational constant,
$\rho_{\mathrm{pl}}$ is the mean density of the planet, and $K$ is the 
fractional
gravitational potential difference between the Roche lobe height and the
planetary surface \citep{Erkaev2007}. In the case of \cvb,
we computed an exact value of $2.068\times 10^{-3}$ for $K$, and
we chose $\eta = 0.15$ following \citet{Salz2015}.
With these parameters, we obtained an energy-limited mass-loss rate of $\dot{M}
= 3.8 \times 10^{15}$~g\,s$^{-1}$ or $0.063$~\mj\myr$^{-1}$, which means  that
the planetary candidate \cvb\ would evaporate within 62~\myr.

Our estimate of the planetary mass-loss rate may be compared to the values derived by \citet{JohnsKrull2016}, based on
their analysis of the H$\alpha$ line profile. \citet{JohnsKrull2016} consider a scenario of
planet-fed accretion and roughly estimated that a mass accretion rate of $3\times 10^{-10}$~M$_{\odot}$\,yr$^{-1}$ or
$2\times 10^{16}$~g\,s$^{-1}$ is required to explain the observations. While this rate is about
an order of magnitude above our estimate, both numbers are subject to significant uncertainty. Moreover,
planetary mass loss may be intermittently enhanced by flaring, and accretion may be episodic as well, as already
suggested by \citet{JohnsKrull2016}.

Flaring is expected to temporarily increase the planetary mass-loss rate.
In their simulations, \citet{Bisikalo2018} find that the mass-loss rate approximately doubles for a ten-fold increase
in the XUV irradiation level. During the \cv\ flare observed by \xmm, the {average} X-ray luminosity increased by about a
factor of five so that we may expect an elevation of about a factor of two in the mass-loss rate.
The mass-loss rate of \cvb\ and its temporal variability certainly remain 
subject to considerable uncertainty due to the uncertain properties of the
planet and the mass-loss process. We therefore conclude that our estimate is not necessarily in contradiction with that
derived by \citet{JohnsKrull2016}.
Mass loss is expected to affect the evolution of a planet like \cvb, which may eventually be stripped down to
a rocky core, as   has been suggested in the case of CoRoT-7\,b \citep{Jackson2010}.
Nevertheless, the estimated lifetime of $62$~\myr\ is about 20 times longer than the system age,
so that we are confident that evaporation does not render a planet at the position of \cvb\ implausible.

\section{Summary and conclusion}

We analyzed one ROSAT, one \xmm,\ and two shorter \chan\ X-ray pointings, which serendipitously
covered \cv.  According to our analysis,
the quiescent X-ray luminosity of \cv\ is about \mbox{$\approx 8\times 10^{29}$~\ergs} and the
X-ray spectrum is consistent with only interstellar absorption.
The X-ray properties of \cv\ are  typical of a WTTS  of similar
spectral type and age.

During the \xmm\ observation, \cv\ showed an X-ray flare that released about $6.8\times 10^{34}$~erg in
the $0.3-9$~keV band. Based on the ephemerides published by \citet{Eyken2012}, \citet{Yu2015}, and \citet{Raetz2016}
it appears plausible that an optical fading event took place during the flare; however,
no simultaneous optical photometry is available to verify this supposition. The X-ray light curve
shows no conclusive evidence for an X-ray counterpart of an optical fading event.    

On the assumption that a hot Jupiter \cvb\ exists, we estimate a mass-loss rate
of $3.8\times 10^{15}$~g\,s$^{-1}$ based on energy-limited escape, which 
yields an approximate lifetime of $62$~\myr, broadly consistent with the value of $20$~\myr\ estimated by \citet{JohnsKrull2016}.
While the true status of the candidate planet \cvb\ cannot be decided based on our
study, the current system age of $2.6$~\myr\ is consistent with
such a configuration.

\begin{acknowledgements}
Based on observations obtained with XMM-Newton, an ESA science mission
with instruments and contributions directly funded by
ESA Member States and NASA.
The scientific results reported in this article are based in
part on data obtained from the Chandra Data
Archive. We have made use of the ROSAT Data Archive of the Max-Planck-Institut für extraterrestrische
Physik (MPE) at Garching, Germany.
SC acknowledges support through DFG projects SCH 1382/2-1 and SCHM 1032/66-1.
\end{acknowledgements} 

\bibliographystyle{aa}
\bibliography{doc.bib}

\end{document}